\documentclass[journal=nalefd,manuscript=letter,layout=twocolumn]{achemso}

\usepackage{chemformula} 
\usepackage[T1]{fontenc} 
\usepackage{dcolumn}   
\usepackage{bm}        
\usepackage{amssymb}   
\usepackage{hyperref}
\usepackage{amsmath}
\usepackage{fixmath}
\usepackage{mathtools}
\usepackage{empheq}
\usepackage[makeroom]{cancel}

\newcommand{\citeasnoun}[1]{Ref.~\citenum{#1}}
\newcommand{\eq}[1]{eq~\ref{#1}}
\newcommand{\fig}[1]{Figure~\ref{#1}}
\newcommand{\SI}[1]{[SI-#1]}

\DeclarePairedDelimiter\norm{\lVert}{\rVert}
\newcommand{\newnorm}[1]{{\left\vert\kern-0.25ex\left\vert\kern-0.25ex\left\vert #1 \right\vert\kern-0.25ex\right\vert\kern-0.25ex\right\vert}}
\newcommand{\vect}[1]{\mathbf{#1}}
\newcommand{\vectgreek}[1]{\mathbold{#1}}

\title{Limits to surface-enhanced Raman scattering near arbitrary-shape scatterers}
\author{J\'{e}r\^{o}me Michon}
\email{jmichon@mit.edu}
\altaffiliation{Contributed equally to this work}
\affiliation{Department of Materials Science and Engineering, Massachusetts Institute of Technology, Cambridge MA, USA}
\author{Mohammed Benzaouia}
\altaffiliation{Contributed equally to this work}
\author{Wenjie Yao}
\affiliation{Department of Electrical Engineering and Computer Science, Massachusetts Institute of Technology, Cambridge MA, USA}
\author{Owen D. Miller}
\affiliation{Department of Applied Physics, Yale University, New Haven CT, USA}
\author{Steven G. Johnson}
\affiliation{Department of Mathematics, Massachusetts Institute of Technology, Cambridge MA, USA}


\begin{document}

\begin{abstract}
	
The low efficiency of Raman spectroscopy can be overcome by placing the active molecules in the vicinity of scatterers, typically rough surfaces or nanostructures with various shapes. This surface-enhanced Raman scattering (SERS) leads to substantial enhancement that depends on the scatterer that is used. In this work, we find fundamental upper bounds on the Raman enhancement for arbitrary-shaped scatterers, depending only on its material constants and the separation distance from the molecule. According to our metric, silver is optimal in visible wavelengths while aluminum is better in the near-UV region. Our general analytical bound scales as the volume of the scatterer and the inverse sixth power of the distance to the active molecule. Numerical computations show that simple geometries fall short of the bounds, suggesting further design opportunities for future improvement. For periodic scatterers, we use two formulations to discover different bounds, and the tighter of the two always must apply. Comparing these bounds suggests an optimal period depending on the volume of the scatterer. 

\end{abstract}

\allowdisplaybreaks

\begin{figure}
	\centering
	\includegraphics[width=\columnwidth, keepaspectratio]{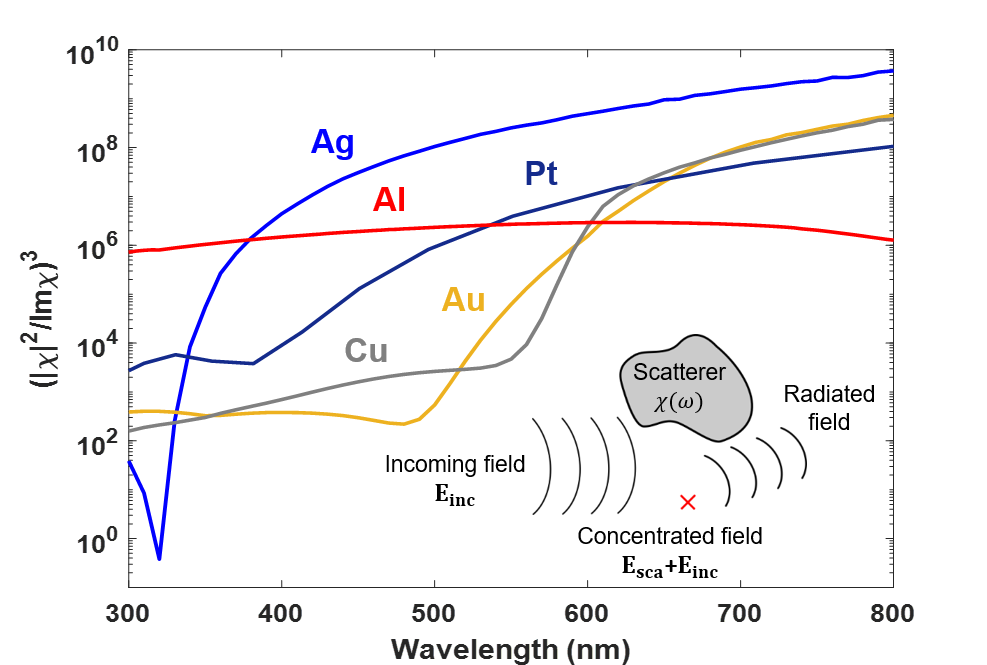}
	\caption{Comparison of the metric $(|\chi|^2/\text{Im}\chi)^3$ for conventional metals used in SERS \cite{Palik1998}. Inset: Schematics of the SERS configuration under study. The pump field is incident onto a scatterer, near which lies the Raman-active chemical. Upon interaction with the pump field, thus material behaves as a dipole emitting a Raman-shifted field. The Raman field interacts with the scatterer and is emitted to the far-field.}
	\label{materials_metric}
\end{figure}

In this Letter, we derive upper limits to surface-enhanced Raman scattering (SERS) \cite{Long1977a,Moskovits1985, Turrell1996,Kneipp1997,Nie1997,Haynes2015,Stiles2008} for arbitrary shapes, both periodic and aperiodic, given only the materials, extending earlier bounds \cite{Miller2016} on linear light emission to a nonlinear process formed from a composition of scattering problems (inset of \fig{materials_metric}), and we show that existing designs such as bowtie antennas are typically far from the theoretical optimum. Earlier work showed that the efficiency of a single light emitter (the local density of states, LDOS) scaled as $|\chi|^2/\text{Im}\chi$ for a material with susceptibility $\chi$ ($=\epsilon-1$) \cite{Miller2016}, but we find that the Raman bounds scale as the \emph{cube} of this (\eq{raman_volume_general}) because they result from nonlinear composition of a light \emph{concentration} bound (in which an incident planewave is concentrated on the Raman molecule) and a light \emph{emission} bound similar to the previous LDOS bounds. The concentration part of our bound ($\sim|\chi|^4/\left(\text{Im}\chi\right)^2$) may also be applicable to many other problems involving light focusing \cite{Vellekoop2007,Mosk2012}. For periodic surfaces, one can gain an additional enhancement to concentration from the contribution of other periods, but we show that there is a trade-off and that the largest benefits (for a single Raman molecule) seem to arise from optimizing individual scatterers. We obtain both analytical formulas within general design regions as well as semi-analytical bounds involving numerical integration for more specific spatial configurations, and we compare typical structures to these bounds. For structures constrained to lie within a subwavelength spherical volume, we show that spherical particles are nearly optimal for certain frequencies. For structures that are allowed to extend into larger volumes, we find that simple geometries such as bowtie antennas \cite{Zhang2015a, Kuhler2014, Dodson2013} are far from our upper limits, suggesting exciting opportunities for improvement in future designs.

SERS was developed to overcome the low efficiency of conventional Raman spectroscopy, as the very small Raman cross-section of most chemicals results in Raman radiation typically on the order of 0.001\% of the power of the pump signal \cite{Long1977a}. In SERS, the chemicals of interest are placed in the vicinity of a scatterer, typically a surface or collection of nanoparticles, which acts as an antenna that both concentrates the incoming pump field at the Raman material's location and enhances the radiated Stokes signal emitted by the Raman material \cite{Kneipp2006,Campion1998}, thereby increasing the collected signal. Charge-transfer mechanisms also lead to a chemical enhancement, although their contribution is smaller than the electromagnetic enhancement effect \cite{Jensen2008, Fromm2006}. Many different materials and antenna geometries have been used for SERS measurements: metals such as silver, gold, or copper, and dielectrics such as silicon carbide or indium tin oxide, were implemented in various shapes such as spheres, triangular prisms, or disks. Several studies have optimized SERS substrates over one or two parameters \cite{Camden2008,LeRu2007,Hao2004,Genov2004,Sundaramurthy2005}. Others have used topology optimization yet only to optimize the concentration of the incident field \cite{Deng2016,Christiansen2019}. Efficiencies up to 12 orders of magnitude larger than that of traditional Raman spectroscopy have been demonstrated, allowing for detection levels down to the single molecule \cite{Nie1997,Kneipp1997} and opening up applications in the fields of biochemistry, forensics, food safety, threat detection, and medical diagnostics. 

However, to the best of our knowledge, no study thus far has looked at the possibility of an upper limit to the enhancement achievable in SERS, and it is therefore not known whether current SERS substrates possess much room for improvement. To investigate the existence of such a bound, a key point is to notice that the process can be decomposed into two linear problems\cite{le2008principles}: concentration of the incident field on the molecule and a dipole emission at the Raman-shifted frequency, as is described in more details below. Upper bounds on the power radiated by a dipole near a scatterer of arbitrary shape were already obtained in \citeasnoun{Miller2016}. Given only the material $\chi$, this is an upper limit for the LDOS for any possible geometric shape in a given region of space near the emitter. The bounding method is based on optimizing the quantity of interest under energy-conservation constraints, using the fact that extinction (linear in the induced fields) is larger than absorption (quadratic in the induced fields). This method has been successfully applied to various other problems \cite{miller2015shape,Shim2018}. Here, we apply this method to obtain a bound on local field concentration enhancement, again for any possible shape given only the material and the bounding volume. Combined with the LDOS limit, we then obtain a bound on the Raman enhancement. We also obtain a \emph{second} bound for the concentration problem using reciprocity in the case of a periodic structure (a similar approach was used to derive the Yablonovitch limit for solar cells from LDOS enhancement \cite{Benzaouia2019}). By comparing the two concentration bounds as a function of period, we obtain a tighter bound for periodic structures. 

\subparagraph{Overview of bounds.}

In order to derive a bound on the Raman enhancement, we consider the configuration represented in the inset of \fig{materials_metric}. An incident ``pump'' planewave $\vect{E_{inc}}$ is scattered by the nanostructure, leading to a near-field enhancement. A Raman-active molecule close to the structure then acquires a dipole moment proportional to the enhanced field, $\vect{p} = \vectgreek{\alpha_\mathrm{R}}\vect{E}$ where $\vectgreek{\alpha_\mathrm{R}}$ is the Hermitian (usually real-symmetric) Raman polarizability tensor \cite{le2008principles}. The power radiated by the dipole in the far field at the Raman frequency is our quantity of interest and is given by $P = |\vectgreek{\alpha_\mathrm{R}} \vect{E}|^2P_\vect{p}$ where $P_{\vect{p}}$ is the power radiated by a unit-vector dipole $\vect{\hat{p}}= \vectgreek{\alpha_\mathrm{R}}\vect{E}/|\vectgreek{\alpha_\mathrm{R}}\vect{E}|$. $P_{\vect{p}}$ can be related to the (electric) LDOS through $\rho_{\vect{p}}=\frac{2\epsilon_0n_b^2}{\pi\omega^2}P_{\vect{p}}$ where $n_b$ is the index of the background medium \cite{Novotny2009}. We note that typically LDOS is defined as the sum of $\rho_{\vect{p}}$ over three orthogonal directions, so that in our notation, the background LDOS is equal to $\rho_b=n_b^3\omega^2/6\pi^2c^3$ \cite{Joulain2005}. The Raman enhancement (compared to the background) is then: 
\begin{equation} q = \frac{|\vectgreek{\alpha_\mathrm{R}}\vect{E}|^2\rho_{\vect{p}}}{\norm{\vectgreek{\alpha_\mathrm{R}}}^2|\vect{E_{inc}}|^2\rho_{b}}\,,\label{radiated_power}\end{equation} 
where $\norm{.}$ is the induced norm (which gives an upper bound on the magnitude of $\vect{\hat{p}}$ for any $\vect{E}$ orientation \cite{lancaster1985theory}). We see that the enhancement comes from two parts: LDOS enhancement ($q_\mathrm{rad}=\rho_\vect{p}/\rho_b$) and local field enhancement ($q_\mathrm{loc}=|\vectgreek{\alpha_\mathrm{R}}\vect{E}|^2/\|\vectgreek{\alpha_\mathrm{R}}\|^2|\vect{E_{inc}}|^2$). To bound the total efficiency, we need to bound both contributions.

\subparagraph{LDOS enhancement.}  

A bound on LDOS enhancement due to scattering by lossy structures can be obtained starting from a result of \citeasnoun{Miller2016}: the maximum LDOS enhancement near a scatterer with susceptibility $\chi$, in a background with Green’s function $\vect{G}$, is given by:
\begin{equation}\label{ldos_bound} q_{\text{rad}}=\frac{\rho_\vect{p}}{\rho_b} \leq 1+\frac{3\pi n_b^2}{2k^3} \frac{|\chi|^2}{\text{Im} \chi}\int_V|\vect{G}\vect{\hat{p}}|^2 \,, \end{equation}  
where the integration is carried over the volume of the scatterer and $k$ the wavenumber in the background medium. Since the direction of $\vect{p}=\vectgreek{\alpha_\mathrm{R}}\vect{E}$ is also to be optimized, we can then obtain a bound using: 
\begin{equation} \begin{split}
\int_V|\vect{G}\vect{\hat{p}}|^2 &= \frac{(\vectgreek{\alpha_\mathrm{R}}\vect{E})^\dagger}{|\vectgreek{\alpha_\mathrm{R}}\vect{E}|} \left(\int_V\vect{G}^\dagger\vect{G}\right)\frac{\vectgreek{\alpha_\mathrm{R}}\vect{E}}{|\vectgreek{\alpha_\mathrm{R}}\vect{E}|} \\
& \leq \newnorm{\vect{G}\vect{U_\mathrm{R}}}^2 \vcentcolon= \norm*{\int_V\vect{U_\mathrm{R}}^\dagger\vect{G}^\dagger\vect{G}\vect{U_\mathrm{R}}} \,,
\end{split}\label{ldos_bound_complement}\end{equation} 
where the columns of $\vect{U_\mathrm{R}}$ -- a $3\times\text{rank}(\vectgreek{\alpha_\mathrm{R}})$ matrix -- are the orthonormal principle axes of $\vectgreek{\alpha_\mathrm{R}}$ with non-zero Raman polarizability. If $\vectgreek{\alpha_\mathrm{R}}$ is invertible we simply obtain $\newnorm{\vect{G}}^2$. On the other hand, if the Raman polarization is along a fixed axis $\vect{\hat{p}}$ we obtain $\newnorm{\vect{G}\vect{\hat{p}}}^2$. 

\subparagraph{Local field enhancement.}
We now obtain a bound on local field concentration by using the same method as in \citeasnoun{Miller2016} while working with a suitable figure of merit. This bound applies to scatterers of \emph{any} shape and scales with the volume of the scatterer. (The focusing of thin parabolic mirrors and lenses provides an example of concentration scaling with volume.) For periodic structures, since there is finite power incident on each unit cell (and a periodic set of foci), we obtain a second bound that scales as the unit-cell area.

\subparagraph{Single-point focusing (volume-scaling bound).}

Let $\vect{x_0}$ be the position of the Raman-active molecule and $\vect{E_{sca}}$ be the scattered field. For \eq{radiated_power}, we want to bound $|\vectgreek{\alpha_\mathrm{R}}\vect{E}(\vect{x_0})|^2$ where $\vect{E}=\vect{E_{sca}}+\vect{E_{inc}}$ is the total field. Recall that the scattered field is given by $\vect{E_{sca}}(\vect{x_0}) = \int_V \vect{G(\vect{x_0},\vect{x})}\vect{P}$ where $\vect{P}=\chi \vect{E}$ is the polarization current \cite{chew1995waves}. 
As explained in \citeasnoun{Miller2016}, the fields are subject to:
\begin{equation}\label{opt-thrm} \text{Im} \int_V \vect{E}^\dagger \vect{P} \leq \text{Im}\int_V\vect{E^\dagger_{inc}} \vect{P} \,, \end{equation}  
which simply states that absorption is smaller than extinction.
For a given unit vector $\vect{\hat{e}}$, maximizing $|\vect{\hat{e}}^\dagger \vectgreek{\alpha_\mathrm{R}}\vect{E_{sca}}(\vect{x_0})|^2$ under the constraint (\ref{opt-thrm}) is equivalent to:
\begin{equation} \max\limits_{\vect{P}}\; \langle \vect{P},\vect{A}\vect{P}\rangle \;\; \text{subject to}\; \langle \vect{P}, \vect{P} \rangle \leq \alpha \text{Re} \langle \vect{b},\vect{P}\rangle \,, \end{equation}
where $\alpha = |\chi|^2/\text{Im} \chi$, $\mathbf{A}\vect{P} = \langle\vect{a},\vect{P}\rangle \vect{a}$, $\vect{a} = \vect{G}^*\vectgreek{\alpha_\mathrm{R}}\vect{\hat{e}}$, $\vect{b} = i\vect{E_{inc}}$ and $\langle\vect{X}, \vect{Y} \rangle = \int_V \vect{X}^\dagger \vect{Y}$. Straightforward variational calculus allows us to solve the optimization problem, yielding\cite{kreutz2009complex,Noce06} \SI{1}:
\begin{equation}
|\vect{\hat{e}}^\dagger\vectgreek{\alpha_\mathrm{R}}\vect{E_{sca}}(\vect{x_0})|^2 \leq \frac{|\chi|^4}{\text{Im}^2 \chi} V|\vect{E_{inc}}|^2   \int_V |\vect{G}^*\vectgreek{\alpha_\mathrm{R}}\vect{\hat{e}}|^2 \,. \end{equation}
A bound on the norm of $\vectgreek{\alpha_\mathrm{R}}\vect{E_{sca}}$ is then obtained similarly to the LDOS result. A simpler bound can be obtained using a spectral decomposition $\vectgreek{\alpha_\mathrm{R}}=\vect{U_\mathrm{R}}\vect{d_\mathrm{R}}\vect{U_\mathrm{R}}^\dagger$ where $\vect{d_\mathrm{R}}$ is a $\text{rank}(\alpha_\mathrm{R})\times \text{rank}(\alpha_\mathrm{R})$ diagonal matrix with entries equal to the nonzero eigenvalues of $\alpha_\mathrm{R}$. In particular, for $\vect{\hat{e}}$ in the column space of $\vectgreek{\alpha_\mathrm{R}}$, we have:
\begin{equation}\begin{split}
 &\int_V |\vect{G}^*\vectgreek{\alpha_\mathrm{R}}\vect{\hat{e}}|^2 \leq \newnorm{\vect{G}^*\vectgreek{\alpha_\mathrm{R}\vect{U_\mathrm{R}}}}^2= \newnorm{\vect{G}^*\vect{\vect{U_\mathrm{R}}d_\mathrm{R}}}^2 \\
 &\; \; \; \; \; \leq \newnorm{\vect{G}^*\vect{\vect{U_\mathrm{R}}}}^2\norm{\vect{d_\mathrm{R}}}^2=\newnorm{\vect{G}\vect{U^*_\mathrm{R}}}^2\norm{\vectgreek{\alpha_\mathrm{R}}}^2\,.
\end{split}\end{equation}
We then conclude by the triangle inequality \cite{lancaster1985theory}:
\begin{equation}
q_\mathrm{loc} \leq \left( 1+ \frac{|\chi|^2}{\text{Im} \chi}\newnorm{\vect{G}\vect{U^*_\mathrm{R}}} \sqrt{V} \right)^2 \,.
\label{bound_incident}\end{equation}
For large enhancement ($q_\mathrm{loc}\gg 1$), the bound is simply given by the second term squared and the material's figure of merit for the concentration bound is the \emph{square} of the usual factor $|\chi|^2/\text{Im} \chi$ from previous works\cite{Miller2016}. Essentially, this arises because concentration involves coupling to two electromagnetic waves: the incoming planewave and the dipole field. The usual material's metric thus comes into play two times. This also explains the presence of the volume of the scatterer (from the coupling with the planewave), and of the integral of the Green's function (from the coupling with the dipole). Identical scalings are also found in the exact results for a quasistatic plasmonic sphere \SI{5}.

\subparagraph{Periodic-array focusing (area-scaling bound).}

In practice, wafer-scale microfabrication techniques favor the manufacturing of repeating patterns over large areas rather than single, isolated structures. Moreover, periodic structures may offer increased SERS performances thanks to interference effects. While the previous bound is still valid for periodic structures by using the \emph{periodic} Green's function, we can also use reciprocity to relate the local field enhancement to LDOS enhancement and obtain a bound that scales as the surface area of the unit cell. We consider a ``2d-periodic'' structure with lattice vectors perpendicular to $\vect{\hat{z}}$ and a unit cell with surface area $S$. We consider both the \emph{scattering} problem with an incident wavevector $\vect{k_0}$ and an amplitude $\vect{E_{inc}}$ and the reciprocal \emph{emission} problem formed by a dipole placed at $\vect{x_0}$ with dipole moment $\vect{\hat{p}}=(\vectgreek{\alpha_\mathrm{R}}\vect{\hat{e}})^*/|\vectgreek{\alpha_\mathrm{R}}\vect{\hat{e}}|$ where $\vect{\hat{e}}$ is an arbitrary unit vector. This emission problem is $-\vect{k_{0\parallel}}$ Bloch-periodic and the radiated far-field can then be decomposed into planewaves with Bloch wavevectors $\vect{k_{\text{nm}}}$ and amplitudes $\vect{T^p_{\text{nm}}}$. Using the same method as in \citeasnoun{Benzaouia2019}, we can relate the near field of the scattering problem $\vect{E(\vect{x_0})}$ to the far-field component  $\vect{T^p_{\text{00}}}$ along $-\vect{k_0}$ in the emission problem through \SI{2}:
\begin{equation}
\frac{\vect{\hat{e}}^\dagger\vectgreek{\alpha_\mathrm{R}} \vect{E}(\vect{x_0})}{|\vectgreek{\alpha_\mathrm{R}}\vect{\hat{e}}|} = \vect{\hat{p}} \cdot \vect{E}(\vect{x_0}) = \frac{2i\epsilon_0S\cos\theta}{k} \vect{T^p_{\text{00}}}\cdot \vect{E_{inc}} \,,
\end{equation}  
where $\theta$ is the polar angle ($\cos\theta = \vect{\hat{k}_0}\cdot \vect{\hat{z}}$). Using this relation, we can now bound the amplitude of $\vectgreek{\alpha_\mathrm{R}} \vect{E}$ using:
\begin{align*} \label{ldos_in} 
&\frac{|\vect{\hat{e}}^\dagger\vectgreek{\alpha_\mathrm{R}}\vect{E}(\vect{x_0})|^2}{\norm{\vectgreek{\alpha_\mathrm{R}}}^2|\vect{E_{inc}}|^2} \leq \frac{4S^2\epsilon_0^2\cos^2\theta}{k^2} |\vect{T^p_{\text{00}}}|^2 \\
&\; \; \; \leq \frac{4S^2\epsilon_0^2\cos\theta}{k^2} \sum_{nm}{ |\vect{T^p_{\text{nm}}}|^2\frac{k_{nm,z}}{\omega\mu_0} \frac{\omega\mu_0}{k} }\\
&\; \; \; = \frac{4S\epsilon_0^2\cos\theta}{k^2}\frac{\omega\mu_0}{k} \;  \text{Re}\int_{S_{+\infty}} \vect{E^p}\times\vect{H^{p*}}\cdot \vect{\hat{z}} \; dS \\
&\; \; \; \leq 8S\epsilon_0^2\cos\theta \frac{\omega \mu_0}{k^3}  P_\vect{p} \,, \stepcounter{equation}\tag{\theequation}\end{align*} 
where we recall that $P_{\vect{p}}$ is the total power radiated by the dipole $\vect{\hat{p}}$. The first inequality is based on Cauchy-Schwartz \cite{lancaster1985theory} while the second one states that the power emitted along $-\vect{k_0}$ ($\propto |\vect{T^p_{\text{00}}}|^2$) is smaller than the total power emitted in the $+\vect{\hat{z}}$ direction, which is then smaller than the total radiated power $P_{\vect{p}}$. The inequalities used in \eq{ldos_in} will be tight (equalities) if $\omega$ is smaller than the first-order diffraction frequency (so that all the power is in $\vect{T^p_{\text{00}}}$ \cite{Joannopoulos2008}) and in the absence of radiated field in the opposite direction (the structure should completely ``block'' the unit-cell's surface). Now using the previous LDOS bound (\eq{ldos_bound}--\ref{ldos_bound_complement}), we conclude: 
\begin{equation}
q_\mathrm{loc} \leq \frac{2Sk^2\cos\theta}{3\pi n_b^4} \left[1+\frac{3\pi n_b^2}{2k^3} \frac{|\chi|^2}{\text{Im} \chi}\newnorm{\vect{G^{per}}\vect{U^*_\mathrm{R}}}^2 \right]  \,,
\label{bound_incident_area}
\end{equation}
where $\vect{G^{per}}$ is the free-space \emph{Bloch-periodic} Green's function.

\subparagraph{Raman enhancement.}

The bound for the Raman enhancement $q=q_\mathrm{loc}(\omega_P)q_\mathrm{rad}(\omega_R)$ is now simply obtained by multiplying the previous bounds (\eq{ldos_bound}--\ref{ldos_bound_complement}) and (\eq{bound_incident}):

\begin{equation} \begin{split}
q &\leq \left( 1+ \frac{|\chi_p|^2}{\text{Im} \chi_p}\newnorm{\vect{G}_p \vect{U^*_\mathrm{R}}}\sqrt{V} \right)^2 \\
&\quad \quad \quad \quad \times \left(1+\frac{3\pi n_b^2}{2k_r^3} \frac{|\chi_r|^2}{\text{Im} \chi_r}\newnorm{\vect{G}_r \vect{U_\mathrm{R}}}^2\right)\\
&\approx \frac{3\pi n_b^2}{2k_r^3}\frac{|\chi_r|^2}{\text{Im} \chi_r} \left(\frac{|\chi_p|^2}{\text{Im} \chi_p}\right)^2V\newnorm{\vect{G}_r\vect{U_\mathrm{R}}}^2\newnorm{\vect{G}_p \vect{U^*_\mathrm{R}}}^2 \,,
\label{raman_volume_general}\end{split}\end{equation}
where the subscripts $p$ and $r$ denote the pump and Raman frequencies at which the variables are evaluated. The second expression is obtained in the case of large enhancement. Also recall that \cite{chew1995waves} (with $k_0$ being the free-space wavenumber): 
\begin{equation}\begin{split}
\vect{G}^\dagger\vect{G} =& \left(\frac{k_0^4}{16\pi^2r^2}\right)\left[\left(1-\frac{1}{(kr)^2}+\frac{1}{(kr)^4}\right)\mathbf{1}\right.\\
&\left.+\left(-1+\frac{5}{(kr)^2}+\frac{3}{(kr)^4}\right)\mathbf{\hat{r}\hat{r}^\dagger}\right] \,.
\end{split}\end{equation} 
If we now assume that the tensor $\vectgreek{\alpha_\mathrm{R}}$ is isotropic and consider simple structures enclosing the scatterer and separated from the Raman-active molecule by a \emph{small} distance $d$, we obtain analytical bounds by considering the lowest order term in $d$ and neglecting far-field terms: 
\begin{equation}q \lesssim \frac{3\pi}{2n_b^6}\beta^2\frac{V}{k_r^3d^6} \frac{|\chi_r|^2}{\text{Im} \chi_r} \left(\frac{|\chi_p|^2}{\text{Im} \chi_p}\right)^2 \,,
\label{raman_volume_sphere}
\end{equation}
where $\beta$ is a geometrical factor equal to $1/6\pi$ for a full sphere, $1/12\pi$ for a half-sphere, and $1/32\pi$ for a half-plane \SI{3}. This fundamental limit scales as $V/d^6$ (compared to $1/d^3$ for LDOS). $1/d^6$ is related to both the radiation of the dipole and the coupling to it, while $V$ is due to the planewave coupling. In practice, the Raman frequency shift is small enough so that the bounds do not change much when the expressions are simply evaluated at the same (pump or Raman) frequency. In this case, the bound is simply proportional to $\left(|\chi|^2/\text{Im} \chi\right)^3$. This material figure of merit can be used to compare the optimal performance of different materials and is shown in \fig{materials_metric}. We note that silver (Ag) has the highest bound at visible wavelengths but is outperformed by aluminum (Al) in the near-UV region.  

The bound of \eq{raman_volume_general} is also valid for a periodic structure after substituting the appropriate periodic Green's function, which can be integrated numerically, for the concentration enhancement term (Raman molecules emit incoherently, so the radiation enhancement is not periodic). Near-field coupling from adjacent unit cells causes the periodic Green's function to increase as the period shrinks so that the maximal bound is obtained for the smallest possible period. However, comparison with the area-scaling bound obtained using \eq{bound_incident_area} shows that this bound isn't tight for small periods. For strong scattering and emission, this area-scaling bound is given by: 
\begin{equation} 
q  \lesssim \frac{3\pi S\cos\theta}{2k_pk_r^3}\frac{|\chi_p|^2}{\text{Im} \chi_p}\frac{|\chi_r|^2}{\text{Im} \chi_r}\newnorm{\vect{G}_p^{\mathbf{per}}\vect{U^*_\mathrm{R}}}^2\newnorm{\vect{G}_r\vect{U_\mathrm{R}}}^2 \,.
 \label{raman_area_general}\end{equation}
By neglecting the Raman frequency shift, this new bound is now proportional to $\left(|\chi|^2/\text{Im} \chi\right)^2$ and scales as the surface area of the unit cell instead of the volume of the scatterer. We can actually see that this area-scaling limit is the same as the volume-scaling one (\eq{raman_volume_general}) when using an effective volume equal to: 
\begin{equation} V_{\text{eff}} = \frac{S \cos \theta}{k_pn_b^2} \frac{\text{Im} \chi_p}{|\chi_p|^2} \,. \label{effective_volume} \end{equation}
This area-scaling bound highlights the fact that for a periodic structure, only a fraction of the actual volume of the scatterer (proportional to the projected unit-cell area)  is effectively ``used'' in the scattering. As explained in more detail below, combining the volume-scaling and area-scaling bounds leads to a tighter bound with different behavior as a function of the period.   
  
\subparagraph{Geometric results.}

\begin{figure*}[h]
	\centering
	\includegraphics[width=\textwidth, keepaspectratio]{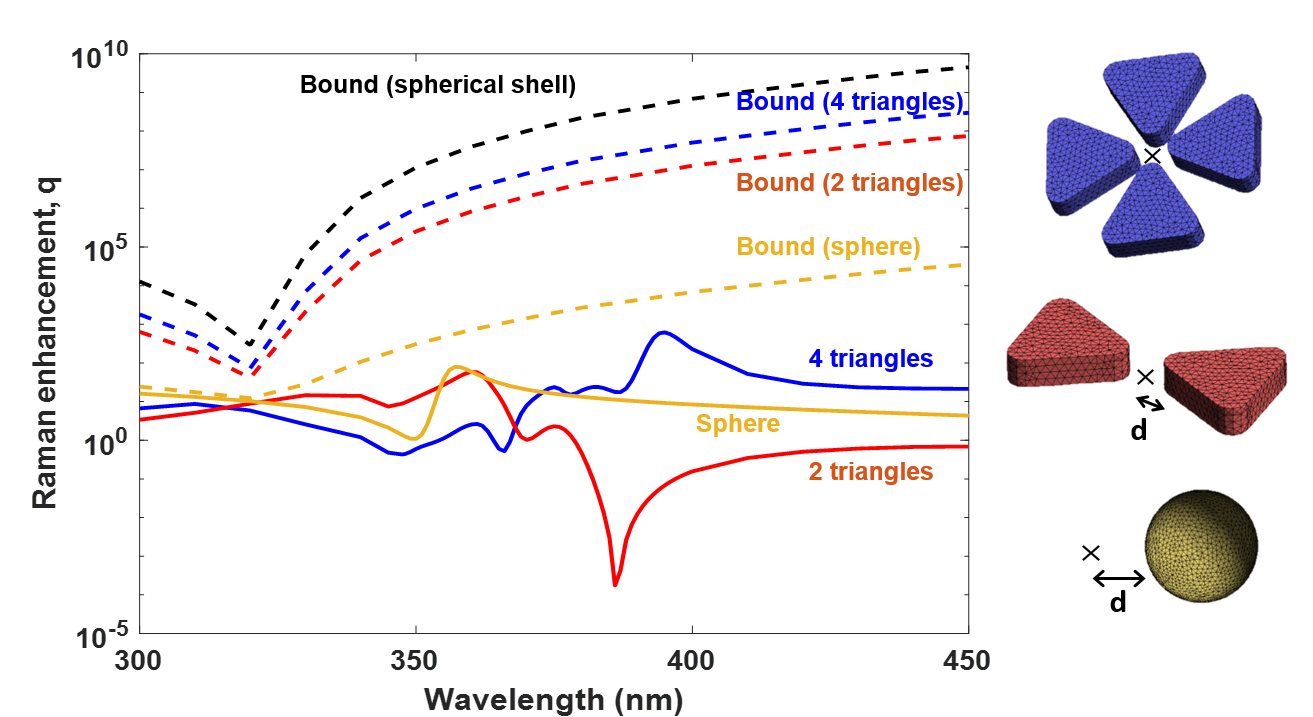}
	\caption{Raman enhancement bounds (dashed lines) and actual performances (full lines) for common SERS Ag structures. The distance to the emitter is $d=20$ nm for all structures. The sphere has a radius of 10 nm. The triangles have a side of 160 nm, height of 30 nm, and tip curvature of 16 nm. The incident field's polarization is aligned with the sphere-emitter and triangle-emitter direction.}
	\label{isolated}
\end{figure*}

The performance of specific structures, assuming an isotropic Raman tensor $\vectgreek{\alpha_\mathrm{R}}$ and a background medium of air, was evaluated using {\sc scuff-em}, an open-source implementation of the boundary-element method \cite{SCUFF1, SCUFF2}. Two simulations were performed for each structure: a scattering simulation to evaluate the field concentration at the Raman material's location, and an emission simulation to evaluate the radiative LDOS \SI{4}. The actual performance of each structure can then be compared to its volume-specific bound by carrying the integration over the volume of the structure (in the expression of $\newnorm{\vect{G}_r}^2$ in \eq{raman_volume_general}), and to a shape-independent bound by carrying the integration over simple geometric structures encompassing the structure (\eq{raman_volume_sphere}). 

\subparagraph{Isolated structures.}

\begin{figure*}[h]
	\centering
	\includegraphics[width=\textwidth, keepaspectratio]{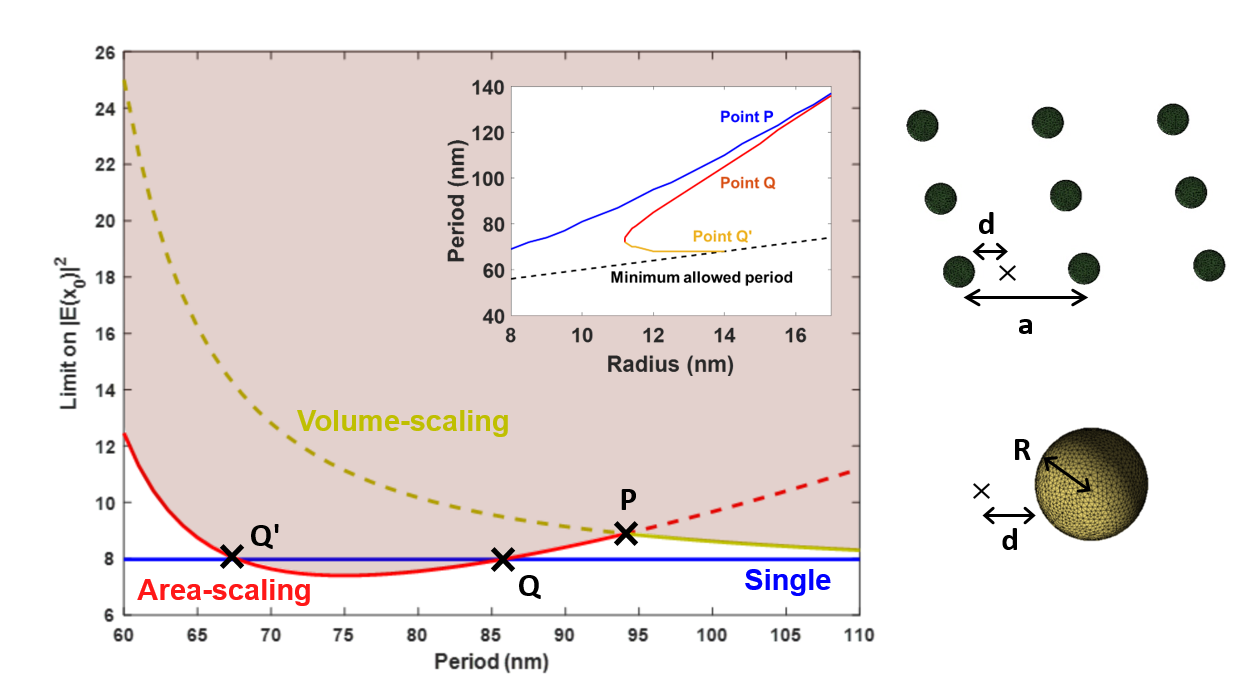}
	\caption{Near-field enhancement bounds for an isolated Ag sphere (bottom right schematics) and a square array of Ag spheres with varying period $a$ (top right schematics). The shaded region indicates forbidden field-concentration values for the periodic case. The spheres have a radius $R = 12$ nm, and the emitter is located $d = 20$ nm away from their surface along the lattice axis. The incident field's polarization is aligned with the sphere-emitter direction and $\lambda = 350$ nm. In the large-enhancement limit, the area-scaling and volume-scaling bounds always intersect at a point, denoted P, where the period equals $\sqrt{k_p[V|\chi|^2/\left[\text{Im}\chi\cos\theta\right]}$. Inset: map of the position of points P, Q, and Q' as a function of sphere radius and lattice period, for $d = 20$ nm.}
	\label{periodic}
\end{figure*}
	
We simulated two of the most common nanostructures used in SERS: triangular prisms used in a bowtie configuration, and a sphere. A sphere has polarizability $\alpha = 3(\epsilon-1)/(\epsilon+2)$ in the electrostatic limit \cite{Bohren1998}, which permits analytical calculation of the concentrated field at resonance ($\textrm{Re}(\chi^{-1})=-1/3$) \SI{4}:
\begin{equation}
|\mathrm{\mathbf{E_{sca}}}|^2 = \dfrac{1}{4\pi^2} \dfrac{1}{(d+R)^6} \left( \dfrac{|\chi|^2}{\mathrm{Im}\,\chi} \right)^2 V^2 |\mathrm{\mathbf{E_{inc}}}|^2 \,,
\end{equation}
with $R$ the radius and $d$ the distance to the emitter. This analytical expression includes all the same factors as our concentration bound (\eq{bound_incident}). By computing $\newnorm{\vect{G}_r}^2$, we find that the sphere's enhancement reaches the bound in the limit $R \ll d \ll \lambda$ \SI{5}. We selected a radius of 10 nm and a distance of 20 nm. All simulated structures were made of silver, which is the best-performing Raman material at visible frequencies (\fig{materials_metric}) and also satisfies the resonance condition for $\chi$, unlike e.g. gold. The geometry of the triangles was taken from \citeasnoun{Kaniber2016a}, with a gap set at 40 nm to readily compare with the sphere results. We included a shape-independent bound by considering the exterior of a spherical shell (entire space minus a sphere of radius $d$), and using the largest volume of all structures, that of the 4-triangle bowtie. The results in \fig{isolated} show that all structures fall short of the shape-independent bound by several orders of magnitude. The performances of bowties also lie far from their shape-specific limits. Only the sphere approaches its bound, at frequencies greater than the plasma frequency of silver.

Finally, it is worth noting that smaller structures get easily closer to the bound compared to larger structures. In the example of the electrostatic sphere, one notices that the ratio of its performance to the bound goes to zero as the radius $R$ increases for a fixed $d$ (Figure SI-2). While both the bound and the actual performance increase with the volume, further shape optimization is required to get closer to the limits for large structures. 
	
\subparagraph{Periodic structures.}
	
To investigate the potential enhancement due to periodicity, we compared the bounds for a single sphere and for a square lattice of similar spheres. We have seen that our Raman limit can be applied to periodic structures by using either of our two bounds on the near-field enhancement: \eq{bound_incident} with the corresponding periodic Green's function, or \eq{bound_incident_area}. We thus only needed to compare the near-field enhancement bounds (\fig{periodic}). The two approaches for the periodic bound yield different geometrical dependencies. The limit of \eq{bound_incident} scales with the volume of the scatterer, kept constant here, and the integral of the periodic Green's function, which decreases towards the non-periodic value as the period increases. The integral of the periodic Green's function also appears in the limit of \eq{bound_incident_area}, yet alongside a factor scaling as the area of the unit cell which reduces the bound for small periods. These behaviors, expected to hold for any scatterer, are indeed observed in \fig{periodic} for arrays of spheres. The actual limit is given by the smaller of the two bounds resulting in different regions in the graph as the period is varied. For periods larger than that of point P (given by \eq{effective_volume} for large enhancement), the volume-scaling bound is limiting because of the reduced interactions between the scatterers of the array. For smaller periods, the performance of the array is limited by the area-scaling bound since the intensity received by each sphere is reduced. Between points Q and Q', this causes the periodic limit to be smaller than the single-sphere limit. Maximum enhancement due to periodicity is still to be found at the smallest possible period, where increased interactions between the scatterers dominate the decrease in incident intensity received by each unit cell.

\subparagraph{Concluding remarks.}

The upper bounds presented in this paper allow a simple estimate of optimal Raman enhancement for arbitrary scatterers. The results show that there is still much room for improvement for large scatterers through further shape optimization. Our analysis of periodic bounds shows the presence of different optimality regions as function of periodicity. While the use of an array can lead to a worse performance for intermediate values of the period, improvement may be still expected for very small periods.

\begin{acknowledgement}

The authors thank Juejun Hu for support and insightful discussions. This work was supported in part by the U.S. Army Research Office through the Institute for Soldier Nanotechnologies under grant W911NF-13-D-0001, and by the National Science Foundation under awards number 1453218 and 1709212. O.D.M. was supported by the Air Force Office of Scientific Research under Grant No. FA9550-17-1-0093.
\end{acknowledgement}

\begin{suppinfo}

Details of calculation to obtain the volume-scaling bound, proof of the reciprocity relation used to derive the area-scaling bound, details of calculation to obtain analytical bounds for simple geometries, intermediate simulation results used to calculate the Raman enhancement of structures, analytical computation of the enhancement for an electrostatic sphere

\end{suppinfo}

\bibliography{Limits_to_Raman}

\newpage
\onecolumn

\begin{center}\Large{\textbf{Supporting information}}\end{center}

\section{1 - Volume-scaling bound}

We want maximize $|\vect{\hat{e}}^\dagger \vectgreek{\alpha_\mathrm{R}}\vect{E_{sca}}(\vect{x_0})|^2 = |(\vectgreek{\alpha_\mathrm{R}}\vect{\hat{e}})^\dagger \vect{E_{sca}}(\vect{x_0})|^2 $ which is equivalent to the following convex quadratic optimization problem: 
\begin{equation} \max\limits_{\vect{P}}\; \langle \vect{P},\vect{A}\vect{P}\rangle \;\; \text{subject to}\; \langle \vect{P}, \vect{P} \rangle \leq \alpha \text{Re} \langle \vect{b},\vect{P}\rangle \,, \label{optimization_prob} \end{equation}
where $\alpha = |\chi|^2/\text{Im} \chi$, $\mathbf{A}\vect{P} = \langle\vect{a},\vect{P}\rangle \vect{a}$, $\vect{a} = \vect{G}^*\vectgreek{\alpha_\mathrm{R}}\vect{\hat{e}}$, $\vect{b} = i\vect{E_{inc}}$ and $\langle\vect{X}, \vect{Y} \rangle = \int_V \vect{X}^\dagger\vect{Y}$.

The optimum of \eq{optimization_prob} must satisy the KKT conditions\cite{kreutz2009complex,Noce06}: \begin{equation}\begin{split} \vect{A}\vect{P} +\lambda (\vect{P}-\frac \alpha 2\vect{b}) &= 0 \,, \\  \langle \vect{P}, \vect{P} \rangle-  \alpha  \text{Re} \langle \vect{b},\vect{P}\rangle &= 0 \,, \end{split}\end{equation}
where $\lambda = -\frac{\langle \vect{a}, \vect{P} \rangle}{\beta}$. The first equation can be written as $\vect{P} = \frac \alpha 2 \vect{b}-\frac 1 \lambda \langle \vect{a}, \vect{P} \rangle \vect{a} = \frac \alpha 2 \vect{b} + \frac \beta 2 \vect{a}$. The second equation then leads to $|\beta| |\vect{a}| = \alpha |\vect{b}|$. Since $\lambda \in \mathbb{R}$, then $\frac{\langle \vect{a},\vect{b}\rangle}{\beta} \in \mathbb{R}$. From this we have $\beta = \pm \alpha \frac{|\vect{b}|\langle \vect{a},\vect{b}\rangle}{|\vect{a}||\langle \vect{a},\vect{b}\rangle|}$. We finally conclude that the optimal value of $\langle \vect{P},\vect{A}\vect{P}\rangle$ is equal to: \begin{equation} \frac{\alpha^2}{4}\left( |\langle \vect{a},\vect{b}\rangle|+|\vect{a}||\vect{b}|\right)^2\leq  \alpha^2 |\vect{a}|^2|\vect{b}|^2 \,. \end{equation}
If we plug in the physical quantities, we get:
\begin{equation} 
|\vect{\hat{e}}^\dagger \vectgreek{\alpha_\mathrm{R}}\vect{E_{sca}}(\vect{x_0})|^2   \leq \left(\frac{|\chi|^2}{\text{Im} \chi}\right)^2  \int_V|\vect{E_{inc}}|^2   \int_V |\vect{G}^*\vectgreek{\alpha_\mathrm{R}}\vect{\hat{e}}|^2 \,.
\end{equation}

\section{2 - Reciprocity relation}

We study a 2d-periodic structure with unit-cell surface area $S$ and ($\vect{b_1},\vect{b_2}$) the reciprocal lattice vectors orthogonal to $\vect{\hat{z}}$\cite{Joannopoulos2008}. We consider the scattering (resp. emission) problem with $\vect{E^s_{inc}} = e^{i\vect{k_0}\cdot \vect{x}} \vect{E_{inc}}$ (resp. $\vect{j} = -i\omega \delta_{\vect{x_0}}\vect{\hat{e}}$, with $-\vect{k_{0\parallel}}$ Bloch boundary-conditions). We can write the outgoing fields in the far field as:
\begin{equation} \vect{E^{e,s}_{out}}= \sum_{n,m} T^{e,s}_{\text{nm}} e^{i\vect{x}\cdot \vect{k^{\text{e,s}}_{\text{nm}}}} \vect{\hat{e}^{\text{e,s}}_{\text{nm}}}, \; \; \vect{H^{e,s}_{out} }= -\sum_{n,m} T^{e,s}_{\text{nm}} e^{i\vect{\vect{x}\cdot k^{\text{e,s}}_{\text{nm}}}}  \frac{\vect{k^{\text{e,s}}_{\text{\text{nm}}}} \times \vect{\hat{e}^{\text{e,s}}_{\text{nm}}}}{\omega \mu_0} \,, \end{equation}
where $|\vect{k^{\text{e,s}}_{\text{nm}}}| = k$, $\vect{k}^{\text{s,e}}_{\parallel}=\pm (\vect{k}_{0\parallel}+n\vect{b_1}+m\vect{b_2})$ and $\vect{\hat{e}^{\text{e,s}}_{\text{nm}}} \cdot \vect{k^{\text{e,s}}_{\text{nm}}}  = 0$. We take $k_{0z}\geq 0$ (with $k_{\text{nm},z} \geq 0$ for $z>0$ and $k_{\text{nm},z}\leq 0$ for $z<0$).\\

From reciprocity, we have: \begin{equation}\label{recip} \int_S(\vect{E^s} \times \vect{H^s}-\vect{E^e}\times\vect{H^s})\cdot \vect{\hat{n}} dS = i\omega  \; \vect{\hat{e}}\cdot \vect{E^{s}(\vect{x_0})} \,, \end{equation}
where $\vect{E^{s}} $ and $\vect{E^{e}} $ are the total fields ($\vect{E^{s}} = \vect{E^{s}_{out}}+\vect{E^{e}_{inc}}$ and $\vect{E^{e}} = \vect{E^{e}_{out}}$).\\
The integration around the lateral boundary is cancelled due to boundary conditions. We now compute the surface integral in the far-field. For $|z|$ large enough, we integrate over the cross section $S_z$:
\begin{equation} \begin{split} \omega \mu_0 \int_{S_{z}} \vect{E^{e}_{out}}\times\vect{H^{s}_{out}}\cdot \vect{\hat{s}} dS &=  -\sum_{n,m,k,l} \int_{S_z} T^{e}_{n,m}T^{s}_{k,l}e^{i\vect{x}\cdot \vect{(k^e_{\text{nm}}+k^s_{\text{kl}})}} \;  \vect{\hat{e}^e_{\text{nm}}}\times (\vect{k^s_{\text{kl}}} \times \vect{\hat{e}^s_{\text{kl}}})\cdot \vect{\hat{z}}\; dS \\
& = S \sum_{nm}T^{e}_{n,m}T^{s}_{n,m}e^{-2ik_{\text{nm},z}z}\left[(\vect{\hat{e}^e_{\text{nm}}} \cdot \vect{\hat{e}^s_{\text{nm}}})\vect{k^s_{\text{nm}}} -(\vect{\hat{e}^e_{\text{nm}}} \cdot \vect{k^s_{\text{nm}}}) \vect{\hat{e}^s_{\text{nm}}} \right]\cdot \vect{\hat{z}}\\
& = S \sum_{nm}T^{e}_{n,m}T^{s}_{n,m}k_{\text{nm},z} e^{-2ik_{\text{nm},z}z} \left[(\vect{\hat{e}^e_{\text{nm}}} \cdot \vect{\hat{e}^s_{\text{nm}}}) -2 (\vect{\hat{e}^s_{\text{nm}}}\cdot \vect{\hat{z}})(\vect{\hat{e}^e_{\text{nm}}} \cdot \vect{\hat{z}}) \right]\\
& = \omega \mu_0 \int_{S_{z}} \vect{E^{e}_{out}}\times\vect{H^{e}_{out}}\cdot \vect{\hat{s}} dS \,.
\end{split}\end{equation}   
The last equality comes from the symmetry of the equation with respect to $e/s$, and the second to last comes from $\vect{k^s_{\text{nm}}} = k_{\text{nm},z}\vect{\hat{z}}-(\vect{k^e_{\text{nm}}}-k_{\text{nm},z}\vect{\hat{z}})$. 

For $z<0$, we also have:
\begin{equation}  \begin{split} \omega \mu_0  \int_{S_{z}} \vect{E^{e}_{out}}\times\vect{H^{s}_{inc}}\cdot \vect{\hat{z}} \; dS  
&= S T^e_{00}[k_{0,z}\vect{\hat{e}^e_{\text{00}}}\cdot \vect{E_{inc}}-\cancel{(\vect{\hat{e}^e_{\text{00}}}\cdot \vect{k_0})}(\vect{E_{inc}}\cdot \vect{\hat{z}})] \\ 
&= ST^e_{00}k_{0,z} \vect{\hat{e}^e_{\text{00}}}\vect{E_{inc}} \,. \end{split}
\end{equation}
Similarly, we find $\int_{S_{z}} \vect{E^{s}_{inc}}\times\vect{H^{e}_{out}}\cdot \vect{\hat{z}} = - \int_{S_{z}} \vect{E^{e}_{out}}\times\vect{H^{s}_{inc}}\cdot \vect{\hat{z}}$. On the other hand, 
\begin{equation}\int_{S_{-z}} \vect{E^{s}_{inc}}\times\vect{H^{e}_{out}}\cdot \vect{\hat{z}} =  \int_{S_{-z}} \vect{E^{e}_{out}}\times\vect{H^{s}_{inc}}\cdot \vect{\hat{z}} =S T^e_{00}k_{0z} \vect{\hat{e}^e_{00}}\cdot [\vect{E_{inc}}-2(\vect{E_{inc}}\cdot \vect{\hat{z}})\vect{\hat{z}}]e^{2ik_{0z}z} \,. \end{equation}

By replacing all integrals in \eq{recip}, we conclude \[ \boxed{\vect{\hat{e}}\cdot \vect{E^{s}}(\vect{x_0}) = \frac{2iS\epsilon_0}{k}\vect{T^e_{\text{00}}}\cdot \vect{E_{inc}}  \cos\theta} \]
where we noted $\vect{T^e_{\text{00}}}=T^e_{00}\vect{\hat{e}^e_{\text{00}}}$ and $\theta$ the incidence angle with respect to $\mathbf{\hat{z}}$. This equation simply relates the field's component $\vect{\hat{e}}\cdot \vect{E^{s}}(\vect{x_0})$ due to an incident plane wave $e^{i\vect{k_0}\cdot \vect{x}} \vect{E_{inc}}$ to the far-field component along $-\vect{k_{0}}$ of the field created by a unit vector dipole $\vect{\hat{e}}$ placed at $\vect{x_0}$ (where the problem in the unit-cell is $-\vect{k_{0\parallel}}$ Bloch-periodic).

\section{3- Induced norm of the integral of the Green's function}

Recall that:
\begin{equation}
\vect{G} = \frac{k_0^2e^{ikr}}{4\pi r}\left[\left(1+\frac{i}{kr}-\frac{1}{(kr)^2}\right)\mathbf{1}+\left(-1-\frac{3i}{kr}+\frac{3}{(kr)^2}\right)\mathbf{\hat{r}\hat{r}^+}\right] \,,
\end{equation} 
so that:
\begin{equation}\begin{split}
\vect{G}^\dagger\vect{G} &= \frac{k_0^4}{16\pi^2r^2}\left[\left(1-\frac{1}{(kr)^2}+\frac{1}{(kr)^4}\right)\mathbf{1}+\left(-1+\frac{5}{(kr)^2}+\frac{3}{(kr)^4}\right)\mathbf{\hat{r}\hat{r}^\dagger}\right] = a(r)\mathbf{1}+b(r)\mathbf{\hat{r}\hat{r}^\dagger} \,.
\end{split}\end{equation} 
If the structure has two mirror symmetry planes orthogonal to $\mathbf{\hat{x}}$, $\mathbf{\hat{y}}$ or $\mathbf{\hat{z}}$, the non-diagonal terms of $\int_V\vect{G}^\dagger\vect{G}$ are zero, and we obtain:
\begin{equation}
\newnorm{\vect{G}}^2 = \norm{\int_V\vect{G}^\dagger\vect{G}} = \int_Va(r) + \max_{j}\int_Vb(r)\frac{x_j^2}{r^2} \,.
\end{equation} 
We can obtain finite analytical expression by integrating over simple geometries and only considering the near-field terms ($\propto 1/r^6$). For spherical shell  of polar angle $\theta$ separated from the Raman molecule by a small distance $d$, we have:
\begin{equation} \int_V \frac{dV}{r^6}= \frac{2\pi(1-\cos\theta)}{3d^3}, \; \int_V \frac{dV}{r^6}\frac{z^2}{r^2}=\frac{2\pi(1-\cos^3\theta)}{9d^3}, \; \int_V \frac{dV}{r^6}\frac{x^2}{r^2}=\frac{\pi(8-9\cos\theta+\cos3\theta)}{36d^3} \,. \end{equation} 
So $\newnorm{\vect{G}}^2n_b^4$ is equal to:
\begin{equation} \frac{1}{24\pi d^3}\left(2-\cos\theta-\cos^3\theta\right) \left[0\leq\theta\leq\frac{\pi}{2}\right],\; \; \; \frac{1}{192\pi d^3}\left(16-17\cos\theta+\cos3\theta\right) \left[\frac{\pi}{2}\leq\theta\leq\pi\right] \,. \end{equation}
For a half-plane, we have:
\begin{equation} \int_V \frac{dV}{r^6}= \frac{\pi}{6d^3}, \; \int_V \frac{dV}{r^6}\frac{z^2}{r^2}=\frac{\pi}{9d^3}, \; \int_V \frac{dV}{r^6}\frac{x^2}{r^2}=\frac{\pi}{36d^3} \,, \end{equation} 
so that $\newnorm{\vect{G}}^2n_b^4 = 1/32\pi d^3$.

\section{4- Concentration and LDOS results for the sphere}

\begin{figure}[H]
	\centering
	\includegraphics[width=\columnwidth, keepaspectratio]{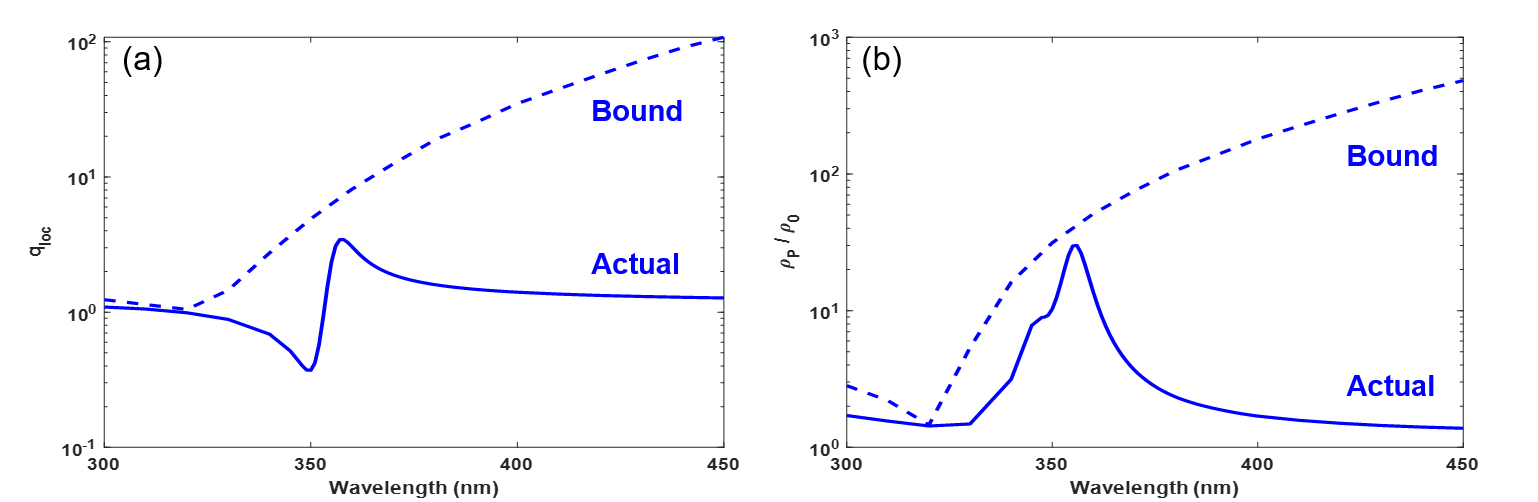}
	\caption{Simulation results and corresponding bounds for a Ag sphere of radius 10 nm and distance to emitter 20 nm. (a) Near-field enhancement with bound from main text. (b) LDOS with bound from \citeasnoun{Miller2016}.}
	\label{sphere_incident}
\end{figure}

\section{5- Concentration enhancement for a plasmonic sphere}

In the quasistatic limit, a plane wave with amplitude $\vect{E_{inc}}$ incident upon a sphere excites a dipole moment: 
\begin{equation}
\vect{p} = \alpha V\vect{E_{inc}} \,,
\end{equation}
where the polarizability $\alpha$ is given by: \begin{equation} 
\alpha = \frac{3(\epsilon-1)}{\epsilon+2} = \frac{1}{1/3+\chi^{-1}} \,.
\end{equation}
On resonance $\text{Re}\chi^{-1} = -\frac 1 3$, such that:
\begin{equation}
\alpha_{max} = \frac{1}{\text{Im}\chi^{-1}}=\frac{|\chi|^2}{\text{Im}\chi} \,.
\end{equation}
The field at a distance $d$ from the sphere of radius $R$ is given by:
\begin{equation}
\vect{E_{sca}} = \frac{1}{4\pi}\left[\frac{3\vect{\hat{n}}(\vect{\hat{n}}\cdot\vect{p})-\vect {p}}{(d+R)^3}\right] \,,
\end{equation}
where $\vect{\hat{n}}$ is the unit vector along the line from the dipole to the measurement point. The amplitude of the field is maximum when $\vect{\hat{n}}$ is along $\vect{p}$ giving:
\begin{equation}
\vect{E_{sca}}=\frac{\vect{p}}{2\pi (d+R)^3} \,.
\end{equation}
Putting all this together, the maximum field concentration outside the sphere at the plasmon frequency is given by:
\begin{equation}
|\vect{E_{sca}}|^2 = \frac{1}{4\pi^2}\frac{1}{(d+R)^6}\left(\frac{|\chi|^2}{\text{Im}\chi}\right)^2V^2|\vect{E_{inc}}|^2 \,.
\label{sphere_analytical}
\end{equation}
This analytical expression includes all the same factors as our bound on the concentrated field, with $\newnorm{\vect{G}_r}^2$ to be compared to the factor $V/4\pi^2(d+R)^6$.
\begin{figure}[h!]
	\centering
	\includegraphics[width=\columnwidth, keepaspectratio]{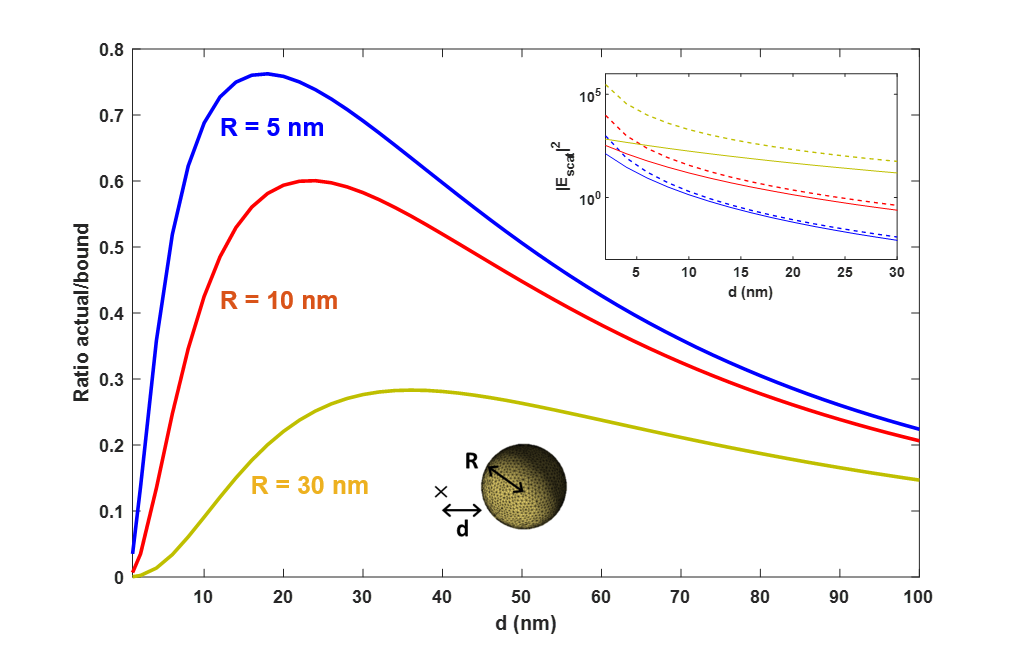}
	\caption{Ratio of the analytical value of of $|\vect{E_{scat}}|^2$ given by \eq{sphere_analytical} to the volume-scaling bound given in the main text, for Ag spheres of different radii at the resonant frequency of Ag. Inset: Comparison of the analytical value (full lines) and the bound (dashed lines) for the same Ag spheres.}
	\label{sphere_ratio}
\end{figure}

We can easily check that the performance of the sphere reaches the bound in the limit of small radius, in particular we can use:
\begin{equation}
\frac{1}{(d+R)^6}\leq\frac{1}{r^6}\leq\frac{1}{(d-R)^6}, \; \frac{x^2}{r^2}\leq\frac{R^2}{(d-R)^2},\; \frac{y^2}{r^2}\leq\frac{R^2}{(d-R)^2}, \; \frac{(d-R)^2}{(d+R)^2}\leq\frac{z^2}{r^2}\leq 1 \,,
\end{equation}
where $z$ is the coordinate along the axis relating the sphere's center and the molecule. We then obtain:
\begin{equation}
\frac 1 V\int_V\frac{dV}{r^6} \xrightarrow[R\rightarrow 0]{} \frac{1}{d^6},\; \frac 1 V\int_V\frac{x^2dV}{r^8} \xrightarrow[R\rightarrow 0]{} 0,\; \frac 1 V\int_V\frac{y^2dV}{r^8} \xrightarrow[R\rightarrow 0]{}0,\; \frac 1 V\int_V\frac{z^2dV}{r^8} \xrightarrow[R\rightarrow 0]{} \frac{1}{d^6} \,.
\end{equation}
We then conclude that $\newnorm{\vect{G}_r}^2 \approx V/4\pi^2d^6$ for $R \ll d$ and that the dipole-sphere performance reaches the upper bound in this limit.

\end{document}